\documentclass{PoS}

\title{VLBI surveys}

\ShortTitle{VLBI surveys}

\author{\speaker{S\'andor Frey}%
         \thanks{This work was supported by the Hungarian Scientific Research Fund (OTKA T046087).}\\
        F\"OMI Satellite Geodetic Observatory, Budapest, Hungary\\
	MTA Research Group for Physical Geodesy and Geodynamics, Budapest, Hungary\\
        E-mail: \email{frey@sgo.fomi.hu}}

\abstract{Systematic surveys of astronomical objects often lead to discoveries, but always provide invaluable information for statistical studies of well-defined samples. They also promote follow-up investigations of individual objects or classes. Surveys using a yet unexplored observing wavelength, a novel technique or a new instrument are of special importance. Significantly improved observing parameters (e.g. sensitivity, angular resolution, monitoring capability) provide new insight into the morphological and physical properties of the objects studied. I give a brief overview of the important Very Long Baseline Interferometry (VLBI) imaging surveys conducted in the past. A list of surveys guides us through the developments up until the present days. I also attempt to show directions for the near future.
}

\FullConference{8th European VLBI Network Symposium\\
		 September 26-29, 2006\\
		 Toru\'n, Poland}

\begin{document}

\section{Introduction}
Astronomical surveys are {\em systematic} studies of the sky (or often only a part of it) in order to {\em explore} an unknown region in the observing {\em parameter space}. Among the possible observing parameters one can mention the observing wavelength or band, the sky coverage, the temporal sampling, the instrument used, its sensitivity, angular resolution, spectral resolution, etc. Due to their completeness, the survey samples are essential for statistical studies of the astronomical objects. In fact, astronomy and astrophysics can be considered as statistical disciplines \cite{long93}. Even the individual objects we understand by comparing them with a class of similar objects. Obviously, larger samples allow us to draw statistically more meaningful conclusions. This is the driving force behind the new surveys, of which the {\em discovery potential} \cite{beck93} essentially depends on the ``size'' of the new parameter space they open up.

A simplistic view of (observational) astronomy would lead us to define three basic categories of astronomers: {\em (1)} who devote their entire lives for studying one or a few individual objects, {\em (2)} who are interested in a well-defined class of objects, and {\em (3)} who want to understand how the Universe as a whole works based on observations of many objects. (Note that the order above by no means reflects the order of the author's preference. I believe we all are mixtures of these three types of astronomers, at different proportions.) The importance of surveys lies in the fact that they provide essential ``raw material'' for all types of astronomical research.

\section{Early VLBI surveys}
Surveys with new observing instruments or techniques are extremely valuable. The first successful Very Long Baseline Interferometry (VLBI) experiments have been conducted at the end of the 1960's (see \cite{mora98} for a review).
The angular resolution provided by VLBI was 3-4 orders of magnitude better than available with any other technique. VLBI made it possible to observe the close vicinity of the central engine of the bright radio-loud active galactic nuclei (AGNs).
Just forty years later, in the present era of modern VLBI, it may seem peculiar that the first sizeable VLBI imaging survey was published only in 1988 \cite{pear88}\footnote{It should be noted, however, that the ``art of the VLBI image reconstruction'', the hybrid mapping method started to develop from the end of the 1970's \cite{read78}.}.
This fundamental work is now referred to as the Pearson--Readhead (PR) survey. The project began in 1977, the first results appeared in 1981 \cite{pear81}. The main goal of the PR survey was to image a flux density-limited source sample ($S>1.3$~Jy at 5~GHz) at northern declinations ($\delta>35^\circ$) avoiding the Galactic plane ($|b|>10^\circ$), in order to support statistical studies of morphologies and superluminal motion. Among the 65 sources selected, 46 were detected with VLBI that time. The fundamentals of the morphological classification of the compact radio emission regions (asymmetric core--jet structures, compact symmetric objects, etc.) were laid down by this work.

Although this review deals with imaging surveys, it is worth mentioning here that there were single-baseline VLBI surveys of considerably larger ($\sim1000$) samples \cite{pres85,mora86}, giving valuable data on radio source compactness and providing initial information for later imaging studies.

The aim of the first Caltech--Jodrell Bank VLBI survey (CJ1, \cite{pola95,thak95,xu95}) -- the first one in the series of ``descendants'' of the PR survey -- was to lower the flux density limit of the PR sample to $S\ge0.7$~Jy at 5~GHz. This led to 135 additional sources, for which snapshot VLBI images were obtained at 1.6 and 5~GHz. The sample became promisingly large to facilitate e.g. studies of cosmological source evolution, misalignment between the pc- and kpc-scale radio structures, and their use as cosmological ``standard objects'' to probe the geometry of the Universe.
The second Caltech--Jodrell Bank survey (CJ2, \cite{tayl94,hens95}) further extended the flux density range ($S\ge0.35$~Jy) while selecting sources based on their flat radio spectra.

By the middle of the 1990's, several hundred sources were imaged in flux density-limited samples, typically at 1.6 and 5~GHz. However, the overall time variability of their compact radio structures was poorly explored. Little was known about their polarization properties in general. Large VLBI survey experiments were clearly constrained by the technical possibilities (frequencies, sensitivity, availability) of the instruments.
The Caltech--Jodrell Bank Flat-spectrum sample (CJF, \cite{tayl96}) of nearly 300 sources provided a homogeneous integration of the survey data available by that time. In fact, the last few observations that made the CJF sample complete were already collected by a new, dedicated VLBI array. That marked the beginning of the ``golden era'' of VLBI surveys in which we still live.

\section{The ``survey machine''}
The Very Long Baseline Array (VLBA) of the U.S. National Radio Astronomy Observatory (NRAO) is an ideal instrument for conducting surveys. First of all, the array of ten antennas is a {\em dedicated} VLBI instrument providing sufficient observing time in practically continuous operation mode \cite{walk95}. On a longer term, this makes the time domain more easily accessible: not only the structures of AGNs, but their evolution can be studied as well. The snapshot capability, the relative ease of polarization observations, the user-friendliness of the VLBA, and the possibility of a highly automated data processing led to a new generation of surveys \cite{list06} some of which we will briefly review here.

The VLBA 2-cm Survey \cite{kell98,zens02,kell04,kova05} observed more than 200 sources at 15~GHz at multiple epochs from 1994 to 2002. The goal was to build up a sample for studying the bulk relativistic {\em motion} in AGN jets. Statistical analysis of the source kinematics showed that the jet motions are practically always directed outwards from the radio core. In most cases, the jet flow seems to follow a continuous path. Evidences for non-ballistic motion were also found. From the distribution of the apparent speeds, it is clear that very high Lorentz factors ($\gamma>25$) are quite rare in AGN jets. The continuation of the 2-cm Survey is the Monitoring of Jets in Active Galactic Nuclei with VLBA Experiments (MOJAVE) program \cite{list05,homa06}. A major improvement is to obtain full linear and circular {\em polarization} images, as well as simultaneous total flux density measurements for the sources in an extended sample.

The VLBA Imaging and Polarimetry Survey (VIPS) \cite{tayl05} is a project to obtain full polarization images of over 1000 AGNs at 5 and 15~GHz. The large and high-quality set of observations will be combined with surveys at other wavelengths: the sky coverage matches that of the Sloan Digital Sky Survey (SDSS) in the optical. The VIPS sources will largely overlap the $\gamma$-ray AGNs to be observed with the forthcoming Gamma-Ray Large Area Space Telescope (GLAST) satellite.

Among the largest VLBI surveys, there are projects initially aimed at specific, primarily non-astrophysical investigations. However, these works often supply highly valuable information for astrophysical studies of individual objects as well.
The VLBA Calibrator Surveys (VCS) \cite{beas02,foma03,petr05,petr06,kova06} provide a list and a searchable database of potential compact phase-reference sources in the northern sky, with accurate coordinates and snapshot images at 2.3 and 8.4~GHz. Currently the VCS extensions contain a total of $\sim3000$ AGNs -- practically each compact flat-spectrum radio source above 200~mJy correlated flux density at 8.4~GHz north of $-30^{\circ}$ declination.

The U.S. Naval Observatory (USNO) Radio Reference Frame Image Database (RRFID) \cite{fey96,fey97,fey00,ojha04} contains images of the sources used for establishing the extragalactic reference frame. The sources are regularly observed at the same frequencies as those used for the astrometry (2.3 and 8.4~GHz), in order to study the effect of the milli-arcsecond (mas) scale source structure on the precision of the reference frame. This work will eventually result in selecting the most suitable reference sources and the improvement of the overall accuracy of the VLBI reference frame. More recently, observations at higher frequencies (24 and 43~GHz) have also become available.

The aim of the VLBA Pre-launch Survey (VLBApls) \cite{foma00} for the VLBI Space Observatory Programme (VSOP) was to select suitable sources for future Space VLBI (SVLBI) observations at 5~GHz. To this end, the correlated flux density of the sources on the longest ground-based baselines was measured. The initial sample of nearly 400 was not constrained by any previous VLBI observation. Snapshot images of the sources -- including a few with first-time VLBI imaging observations -- were also made available.

\section{VLBI in space}

The first satellite dedicated to SVLBI was launched in Japan in 1997 \cite{hira00a}, providing space--ground baselines up to 3 times longer than available on the ground only. About a quarter of the HALCA observing time was devoted to the 5-GHz VSOP Survey Program \cite{hira00b,hori04,scot04,love04,dods06}. Although the observations involved limited ground-based resources (typically 2-4 radio telescopes), the unprecedented resolution allowed us to study the sub-mas structure of a flux density-limited sample of bright sources and to obtain higher brightness temperature measurements (or lower limits to the brightness temperatures). The VSOP Survey is a fine example of how a new instrument can play an important role in conducting surveys. The brightness temperature distribution of the sources shows a clear tail well above the $10^{12}$~K inverse-Compton limit confirming the presence of relativistic beaming. The average relative visibilities of the sources indicate that a large fraction (40\%) of the total flux density of the sources originate in very compact ($\le0.2$~mas) components \cite{hori04}.

A sub-sample of 27 sources from the classical PR survey was imaged with high angular resolution and high dynamic range with VSOP \cite{list01a}. New information on the parsec-scale radio structure of this statistically complete core-selected sample of AGNs, together with extensive data taken at various wavelengths allowed studies of the relativistic beaming effects in compact radio sources. Many previously known correlations between different source properties have been confirmed. Several new trends that support the beaming model have been discovered \cite{list01b}. A significant fraction of the sources studied have brightness temperature in excess of $10^{12}$~K. A relationship between the brightness temperature and the intra-day variability type has also been found \cite{ting01}.

\section{Multi-band approach and deep fields}

Astronomical surveys are of course not limited to VLBI. In fact, surveys at present flourish, partly due to the new powerful (often space-based) facilities and because future instruments need target and/or calibrator object lists. Obviously, the astrophysical outcome of a survey is greatly boosted by data taken at multiple wavebands. From a VLBI point of view, among the potential enhancements one can mention the X- and $\gamma$-ray emission from jets, the broad-band spectral energy distribution derived for AGNs, the relativistic beaming indicators like the optical emission line widths, or the cosmological redshifts obtained from optical spectroscopy.

The Deep Extragalactic VLBI-Optical Survey (DEVOS) \cite{moso06} --~currently in its pilot stage~-- attempts to detect one or two orders of magnitude fainter objects in radio than those targeted in large-scale VLBI surveys at present. Although complete above a certain flux density limit, the source samples suffer from a strong luminosity selection: the higher the distance, the larger the luminosity of the AGNs that can enter the sample \cite{gurv04}. To reach conclusive results on the intrinsic source properties, their evolution, and the possible cosmological effects, sources of similar luminosities should be compared across the whole redshift range. The optical identifications and redshifts for DEVOS are provided by the SDSS.

Deep field surveys are necessarily restricted to small parts of the sky, but greatly advance our knowledge about the faint and/or distant source population (see \cite{garr05b} for a review). A stunning example is the Hubble Deep Field North (HDF-N). Highly sensitive wide-field VLBI observations identified sub-mJy sources for which the radio emission is dominated by the AGN process \cite{garr01}. Deep VLBI imaging of the NOAO Bootes field \cite{garr05a} also demonstrated that the wide-field approach can be successfully employed to image many potential targets simultaneously.
This field has also been explored with phase-referenced VLBA observations at 5 GHz \cite{wrob05}.
Compact radio emission at mJy level has been detected in about 40\% of
faint radio sources.

\section{Outlook}
The role of the European VLBI Network (EVN) in carrying out VLBI surveys has so far been less significant than that of the VLBA. The main reason is that the session-based (not continuous) observing and thus the limited time available makes large all-sky surveys non-competitive at the EVN. On the other hand, the rapid technological advances -- the increasing data rates and new telescopes with large collecting area foreseen -- could somewhat compensate for the observing time. The resulting sensitivity of the network could be utilised by deep surveys, concentrating on small regions of the sky. This has already been demonstrated using the the wide-field technique \cite{garr01}.
With the advent and rapid development of e-VLBI \cite{szom06}, the way how the technique works will soon be revolutionised. The EVN and MERLIN could eventually be combined in a single, highly reliable instrument. This would certainly open up new possibilities for surveys as well.

\appendix
\section{VLBI imaging surveys on the web}

One of the important aspects of the surveys is the {\em availability} of their results for the entire astronomical community. Apart from the publications in scientific journals, it is straightforward to provide access to the data through the internet. The major VLBI surveys did a good job with setting up websites from where the basic information on the sample and the observations, the related references, the source images and the model parameters can be obtained. Below, I give an up-to-date list of relevant web addresses for easy reference.

\begin{itemize}
\item Pearson--Readhead and Caltech--Jodrell Bank surveys:\\
  {\tt www.astro.caltech.edu/$\sim$tjp/cj/}
\item VLBA 2-cm Survey:\\
  {\tt www.cv.nrao.edu/2cmsurvey/}
\item Monitoring of Jets in Active Galactic Nuclei with VLBA Experiments (MOJAVE):\\
  {\tt www.physics.purdue.edu/astro/MOJAVE/}
\item USNO Radio Reference Frame Image Database (RRFID):\\
  {\tt rorf.usno.navy.mil/rrfid.shtml}
\item VLBA Calibrator Surveys (VCS):\\
  {\tt www.vlba.nrao.edu/astro/calib/}
\item VSOP Pre-launch Survey (VLBApls):\\
  {\tt www.vlba.nrao.edu/astro/obsprep/sourcelist/6cm/}
\item VSOP Survey:\\
  {\tt www.vsop.isas.jaxa.jp/survey/}
\item VLBI Imaging and Polarimetry Survey (VIPS):\\
  {\tt www.phys.unm.edu/$\sim$gbtaylor/VIPS/}
\end{itemize}


\begin{thebibliography}{99}

\bibitem{beas02}
A.J.~Beasley et al. 2002, {\em The VLBA Calibrator Survey--VCS1}, {\em ApJS} {\bf 141}, 13

\bibitem{beck93}
S.V.W.~Beckwith 1993, {\em Surveys in Astronomy}, ASP Conference Series {\bf 43}, 303

\bibitem{dods06}
R.~Dodson et al. 2006, {\em The VSOP Survey: Final Indvidual Results}, these Proceedings, \pos{PoS(8thEVN)070}

\bibitem{fey97}
A.L.~Fey \& P. Charlot 1997, {\em VLBA Observations of Radio Reference Frame Sources. II. Astrometric Suitability Based on Observed Structure}, {\em ApJS} {\bf 111}, 95

\bibitem{fey00}
A.L.~Fey \& P. Charlot 2000, {\em VLBA Observations of Radio Reference Frame Sources. III. Astrometric Suitability of an Additional 225 Sources}, {\em ApJS} {\bf 128}, 17

\bibitem{fey96}
A.L.~Fey, A.W.~Clegg \& E.B. Fomalont 1996, {\em VLBA Observations of Radio Reference Frame Sources. I.}, {\em ApJS} {\bf 105}, 299

\bibitem{foma00}
E.B.~Fomalont et al. 2000, {\em The VSOP 5~GHz Continuum Survey: The Prelaunch VLBA Observations}, {\em ApJS} {\bf 131}, 95

\bibitem{foma03}
E.B.~Fomalont et al. 2003, {\em The Second VLBA Calibrator Survey: VCS2}, {\em AJ} {\bf 126}, 2562

\bibitem{garr05b}
M.A.~Garrett 2005, {\em Deep field surveys - A radio view}, EAS Publications Series {\bf 15}, 73

\bibitem{garr05a}
M.A.~Garrett, J.M.~Wrobel \& R. Morganti 2005, {\em Deep VLBI Imaging of Faint Radio Sources in the NOAO Bootes Field}, {\em ApJ} {\bf 619}, 105

\bibitem{garr01}
M.A.~Garrett et al. 2001, {\em AGN and starbursts at high redshift: High resolution EVN radio observations of the Hubble Deep Field}, {\em A\&A} {\bf 366}, L5

\bibitem{gurv04}
L.~Gurvits 2004, {\em Surveys of compact extragalactic radio sources}, {\em NewAR} {\bf 48}, 1211

\bibitem{hens95}
D.R.~Henstock et al. 1995, {\em The Second Caltech--Jodrell Bank VLBI Survey. II. Observations of 102 of 193 Sources}, {\em ApJS} {\bf 100}, 1

\bibitem{hira00a}
H.~Hirabayashi et al. 2000a, {\em The VLBI Space Observatory Programme and the Radio-Astronomical Satellite HALCA}, {\em PASJ} {\bf 52}, 955

\bibitem{hira00b}
H.~Hirabayashi et al. 2000b, {\em The VSOP 5 GHz AGN Survey I. Compilation and Observations}, {\em PASJ} {\bf 52}, 997

\bibitem{homa06}
D.C.~Homan \& M.L.~Lister 2006, {\em MOJAVE: Monitoring of Jets in Active Galactic Nuclei with VLBA Experiments. II. First-Epoch 15~GHz Circular Polarization Results}, {\em AJ} {\bf 131}, 1262

\bibitem{hori04}
S.~Horiuchi et al. 2004, {\em The VSOP 5~GHz Active Galactic Nucleus Survey. IV. The Angular Size/Brightness Temperature Distribution}, {\em ApJ} {\bf 616}, 110

\bibitem{kell98}
K.I.~Kellermann et al. 1998, {\em Sub-Milliarcsecond Imaging of Quasars and Active Galactic Nuclei}, {\em AJ} {\bf 115}, 1295

\bibitem{kell04}
K.I.~Kellermann et al. 2004, {\em Sub-Milliarcsecond Imaging of Quasars and Active Galactic Nuclei. III. Kinematics of Parsec-Scale Radio Jets}, {\em ApJ} {\bf 609}, 539

\bibitem{kova05}
Y.Y.~Kovalev et al. 2005, {\em Sub-Milliarcsecond Imaging of Quasars and Active Galactic Nuclei. IV. Fine-Scale Structure}, {\em AJ} {\bf 130}, 2473

\bibitem{kova06}
Y.Y.~Kovalev et al. 2006, {\em The Fifth VLBA Calibrator Survey: VCS5}, {\em AJ}, in press, {\tt astro-ph/0607524}

\bibitem{list06}
M.L.~Lister 2006, {\em Structure and Evolution of Blazar Jets: Recent Results from VLBI Surveys}, ASP Conference Series {\bf 350}, 139

\bibitem{list05}
M.L.~Lister \& D.C.~Homan 2005, {\em MOJAVE: Monitoring of Jets in Active Galactic Nuclei with VLBA Experiments. I. First-Epoch 15~GHz Linear Polarization Images}, {\em AJ} {\bf 130}, 1389

\bibitem{list01b}
M.L.~Lister, S.J.~Tingay \& A.R.~Preston 2001, {\em The Pearson--Readhead Survey of Compact Extragalactic Radio Sources from Space. II. Analysis of Source Properties}, {\em ApJ} {\bf 554}, 964

\bibitem{list01a}
M.L.~Lister et al. 2001, {\em The Pearson--Readhead Survey of Compact Extragalactic Radio Sources from Space. I. The Images}, {\em ApJ} {\bf 554}, 948

\bibitem{long93}
M.S.~Longair 1993, {\em A Survey of Surveys}, ASP Conference Series {\bf 43}, 313

\bibitem{love04}
J.E.J.~Lovell et al. 2004, {\em The VSOP 5 GHz Active Galactic Nucleus Survey. II. Data Calibration and Imaging}, {\em ApJS} {\bf 155}, 27

\bibitem{mora86}
D.D.~Morabito et al. 1986, {\em VLBI observations of 416 extragalactic radio sources}, {\em AJ} {\bf 91}, 1083

\bibitem{mora98}
J.M.~Moran 1998, {\em Thirty Years of VLBI: Early Days, Successes, and Future}, ASP Conference Series {\bf 164}, 1

\bibitem{moso06}
L.~Mosoni et al. 2006, {\em Deep Extragalactic VLBI-Optical Survey (DEVOS) I. Pilot MERLIN and VLBI observations}, {\em A\&A} {\bf 445}, 413

\bibitem{ojha04}
R.~Ojha et al. 2004, {\em VLBI Observations of Southern Hemisphere ICRF Sources. I.}, {\em AJ} {\bf 127}, 3609

\bibitem{pear81}
T.J.~Pearson \& A.C.S.~Readhead 1981, {\em The milli-arcsecond structure of a complete sample of radio sources. I - VLBI maps of seven sources}, {\em ApJ} {\bf 248}, 61

\bibitem{pear88}
T.J.~Pearson \& A.C.S.~Readhead 1988, {\em The milliarcsecond structure of a complete sample of radio sources. II. First-epoch maps at 5 GHz}, {\em ApJ} {\bf 328}, 114

\bibitem{petr05}
L.~Petrov et al. 2005, {\em The Third VLBA Calibrator Survey: VCS3}, {\em AJ} {\bf 129}, 1163

\bibitem{petr06}
L.~Petrov et al. 2006, {\em The Fourth VLBA Calibrator Survey: VCS4}, {\em AJ} {\bf 131}, 1872

\bibitem{pola95}
A.G.~Polatidis et al. 1995, {\em The First Caltech--Jodrell Bank VLBI Survey. I. $\lambda =$~18 Centimeter Observations of 87 Sources}, {\em ApJS} {\bf 98}, 1

\bibitem{pres85}
R.A.~Preston et al. 1985, {\em A VLBI survey at 2.29 GHz}, {\em AJ} {\bf 90}, 1599

\bibitem{read78}
A.C.S.~Readhead \& P.N.~Wilkinson 1978, {\em The mapping of compact radio sources from VLBI data}, {\em ApJ} {\bf 223}, 25

\bibitem{scot04}
W.K.~Scott et al. 2004, {\em The VSOP 5 GHz Active Galactic Nucleus Survey. III. Imaging Results for the First 102 Sources}, {\em ApJS} {\bf 155}, 33

\bibitem{szom06}
A.~Szomoru et al. 2006, {\em Recent e-EVN developments}, these Proceedings, \pos{PoS(8thEVN)053}

\bibitem{tayl94}
G.B.~Taylor et al. 1994, {\em The Second Caltech-Jodrell Bank VLBI Survey. I. Observations of 91 of 193 sources}, {\em ApJS} {\bf 95}, 345

\bibitem{tayl96}
G.B.~Taylor et al. 1996, {\em A Complete Flux-Density-limited VLBI Survey of 293 Flat-Spectrum Radio Sources}, {\em ApJS} {\bf 107}, 37

\bibitem{tayl05}
G.B.~Taylor et al. 2005, {\em VLBA Imaging Polarimetry of Active Galactic Nuclei: An Automated Approach}, {\em ApJS} {\bf 159}, 27

\bibitem{thak95}
D.D.~Thakkar et al. 1995, {\em The First Caltech--Jodrell Bank VLBI Survey. II. $\lambda =$~18 Centimeter Observations of 25 Sources}, {\em ApJS} {\bf 98}, 33

\bibitem{ting01}
S.J.~Tingay et al. 2001, {\em Measuring the Brightness Temperature Distribution of Extragalactic Radio Sources with Space VLBI}, {\em ApJ} {\bf 549}, L55

\bibitem{walk95}
R.C.~Walker 1995, {\em What the VLBA Can Do for You: Capabilities, Sensitivity, Resolution and Image Quality}, ASP Conference Series {\bf 82}, 133

\bibitem{wrob05}
J.M.~Wrobel et al. 2005, {\em Faint Radio Sources in the NOAO Bo\"otes Field:
VLBA Imaging and Optical Identifications}, {\em AJ} {\bf 130}, 923

\bibitem{xu95}
W.~Xu et al. 1995, {\em The First Caltech--Jodrell Bank VLBI Survey. III. VLBI and MERLIN Observations at 5~GHz and VLA Observations at 1.4~GHz}, {\em ApJS} {\bf 99}, 297

\bibitem{zens02}
J.A.~Zensus et al. 2002, {\em Sub-Milliarcsecond Imaging of Quasars and Active Galactic Nuclei. II. Additional Sources}, {\em AJ} {\bf 124}, 662



\end{thebibliography}
\end{document}